\documentclass[9pt,a4paper]{article}
\usepackage{blindtext}
\usepackage[T1]{fontenc}
\usepackage[utf8x]{inputenc}
\usepackage{amsmath}
\usepackage{graphicx}
\graphicspath{{images/}}
\usepackage{subfigure}
\usepackage{hyperref}
\usepackage{authblk}
\usepackage[normalem]{ulem}
\hypersetup{
    colorlinks=true,
    linkcolor=blue,
    filecolor=magenta,      
    urlcolor=cyan,
}
\usepackage{vmargin}
\setmarginsrb{2.5 cm}{1.5 cm}{2.5 cm}{1.5 cm}{0.5 cm}{0.5 cm}{0.5 cm}{0.5 cm}
\date{\today}
\usepackage{setspace}
\usepackage{color,soul}
\title{Model studies on motion of respiratory droplets driven through a face mask}
\author{Rahul Karmakar \footnote{Corresponding author}}
\affil[1]{Department of Chemical, Biological and Macro-Molecular Sciences, S. N. Bose National Centre for Basic Sciences, Block-JD, Sector-III, Salt Lake Kolkata 700106, India. \\
rahul.physics2017@gmail.com}

\author[2]{Aishani Ghosal}

\affil[2]{
 Department of Biomedical Engineering, Tel Aviv University, Tel Aviv 6997801, Israel  \\
 aishanig@mail.tau.ac.il
}%

\author[1]{J Chakrabarti}

\affil[1]{Department of Chemical, Biological and Macro-Molecular Sciences, S. N. Bose National Centre for Basic Sciences, Block-JD, Sector-III, Salt Lake Kolkata 700106, India.\\
jaydeb@bose.res.in}%

\begin{document}
\maketitle


\begin{abstract}

Face masks are used to intercept respiratory droplets to prevent spreading of air-borne diseases.  Designing face masks with better efficiency 
needs microscopic understanding on how respiratory droplets move through a mask. Here we study a simple model on the interception of droplets by a face mask. The mask is treated as a polymeric network in an asymmetric confinement, while the droplet is taken as a micrometer sized tracer colloidal particle, subject to driving force that mimics the breathing.  We study numerically, using the Langevin dynamics, the tracer particle permeation through  the polymeric network. We show that the permeation is an activated process following an Arrhenius dependence on temperature. The potential energy profile responsible for the activation process increases with tracer size, tracer bead interaction, network rigidity and decreases with the driving force and confinement length. A deeper energy barrier led to better efficiency to intercept the tracer particles of a given size in the presence of driving force at room temperature. Our studies may help to design mask with better efficiency.

\end{abstract}

\subsection*{Keywords}
Langevin dynamics simulation, Confined polymer network,   colloidal particles, permeation


Using face mask (FM)  has been commonplace much beyond the usual medical purpose \cite{strasser2020history}, since the COVID-19 pandemic. The World Health Organization (WHO) mandates the use of FM to prevent the spread of COVID-19 \cite{lyu2020community}. By now it is established \cite{zhao2020household,bazant2021guideline,karmacharya2021advances} that the SARS-COV-2  virus responsible for COVID-19 is air-borne, like many other air-borne diseases, namely, tuberculosis, pneumonia and so forth. FM is the simplest measure to prevent spreading of any air-borne disease \cite{bazant2021guideline,poon2020soft,feng2020rational}by intercepting respiratory droplets of micrometer size, emitted when an affected person talks, sneezes or coughs. Respiratory micro-droplet propagation through air  has been intensively studied\cite{riley1978airborne,wells1934air,poon2020soft,netz2020physics,bovzivc2021relative}. However, micro-droplet movement through an FM is far from understood, although this understanding is essential to design FMs with better efficiency\cite{karmacharya2021advances}, breath-ability\cite{ye2022breathable} and re-usability\cite{ye2022breathable,phan2020reusable}.

 FMs are typically made of fabric materials of entangled polymeric networks\cite{zhao2020household}, which allow essential molecules to be exchanged with the atmosphere and the human body but stop the movement of larger particles \cite{zhao2020household,konda2020aerosol}. Normal  N95 FMs with fibres of micrometer range are commercially used. They filter 95$\%$ of the  micro-droplets through electrostatic capture and mechanical interception\cite{ullah2020reusability}. But they show reduction in efficiency in humid conditions due to charge loss\cite{karabulut2021electrospun}. In search for stable FM performances, different possibilities are explored which include, for instance, using tribo-electric generators\cite{karmacharya2021advances,liu2018self,ghatak2021design}, mixture of fabrics with different compositions\cite{konda2020aerosol}, and multiple layers\cite{sharma2021secondary}.
Nano-fibres\cite{ullah2020reusability} and  metal−organic framework filters  \cite{hao2019electrospun,tang2020materials}  are shown to be promising route for better FM. However, they are plagued with breathing difficulty due to pores of nano-meter sizes. Breath-ability is an indicator of resistance to airflow by the mask while breathing and given in terms of the pressure drop inside the mask during airflow \cite{konda2020aerosol}. The average pressure difference in normal breathing is found to be around 2.5 ± 0.4 Pa\cite{konda2020aerosol}. Reduction of leakage through FM is also necessary for better efficiency\cite{verma2020visualizing}. 

Given this backdrop, theoretical understanding of various aspects to control the efficiency of intercepting micro-droplets by FM would be highly imperative. A numerical study using the continuum fluid dynamics equations suggests that the droplet movement is a combined effect of capillary action and breathing force inside the FM\cite{yi2005numerical}. At a more microscopic level, one may consider the respiratory droplets as colloidal particles of micrometer size moving through a porous polymeric network structure. Particle movement through a polymer network has been widely studied \cite{kumar2019transport,singh2020non,cho2020tracer,sorichetti2021dynamics,anderson2021subtle,samanta2016tracer}. However, in case of FM there are a few important aspects: (1) The WHO recommends that mask should be made up of three layers\cite{sharma2021secondary,10665-331693}: a polar material like cotton layer inside towards mouth and hydrophobic material in the outermost layer \cite{karmacharya2021advances,10665-331693,lustig2020effectiveness}. (2) The penetration of the droplet inside the porous mask also depends on the wettability of the fabric materials by the micro-droplets\cite{forouzandeh2021face}. (3) It is necessary to consider the appropriate force needed to ensure the breath-ability condition. 

Considering these aspects, we study a  model to understand micro-droplet movement inside an FM. We treat the micro-droplets as micro-meter sized tracer colloidal particles (TCP). The FM is modelled in terms of a network of polymeric strands with two different kinds of beads of micrometer size. One class of beads (h-beads) are taken to mutually interact more strongly than the other class of beads (p-beads). The h-beads mimic hydrophobic species those are known to collapse\cite{chakrabarti2015analytical,berne2009dewetting,ten2002drying} and ensure strong entanglement of the network. The p-beads interact more strongly with TCPs. The network is under an asymmetric confinement, consisting of walls, each of which is having favourable interaction with a given type of bead. One wall preferentially interacts with h-beads, called the h-wall at $z=0$ and the other one interacts favourably with p-beads denoted as the p-wall at $z=z_{P}$. This serves as a prototype for the three layered model suggested by WHO\cite{10665-331693}. Our attention is specifically on the middle layer of the tri-layer FM. The TCPs move from one wall to the other through a heterogeneous media in presence of pressure difference needed for breath-ability. The TCPs larger than the mean pore size will be stopped due to geometrical constraints. Hence, our focus is on the factors which control the TCP flux slightly below the natural stopping size.  

We study the model using the Langevin dynamics simulations\cite{schneider1978molecular}. Let us consider the movement of the TCPs from h-wall to p-wall. The fraction of tracer particles reaching the p -wall is the permeation $P$ through the mask, while the efficiency of the mask $e=100-P$. We find that the permeation of the TCPs is an activated process, the energy landscape responsible for the activated process being governed by the  TCP and p-bead interaction. Large activation barrier leads to a lower $P$ and hence, larger $e$. We study the efficiency of intercepting the TCPs, fixing size, the driving force and the ambient temperature while varying composition of the polymeric strands, TCP interaction with the p-beads, network rigidity and confinement size. We find that a 50:50 mixture of polymeric beads and stronger TCP and p-bead interaction gives better efficiency. Mask efficiency linearly increases with increasing network rigidity and with decreasing confinement size. These may be helpful to design a mask with large efficiency ensuring normal breath-ability condition.

\textbf{Model and simulation.} - Each polymeric strand is composed of two kinds of beads of the same diameter $\sigma$ and mass $m$ in a given h:p ratio, randomly distributed over the strand. The non-bonded interaction between two monomers is taken through the Lennard-Jones(L-J) 12-6 potential: 
$V_{\alpha \beta}(r_{ij}) = 4\epsilon_{\alpha \beta} [ (\frac{\sigma}{r_{ij}})^{12} - (\frac{\sigma}{r_{ij}})^{6}] $, $r_{ij} < 3\sigma$. Here $\alpha(=h,p)$ and $\beta(=h,p)$ stand for the bead types and $r_{ij}$ is separation between two beads. The bonded interaction corresponding to stretching between two neighbouring beads:$V_{bond} (r_{ij}) = \frac{1}{2} k (r_{ij} - r_{0})^{2}$ 
where $r_{ij}$ is the distance between neighbouring monomers and $r_{0}=1.5\sigma$ is the equilibrium distance between monomers and $k$ the force constant. The change in bond  angle costs energy:
$V_{angle} (\theta) = \frac{1}{2} k_{\theta} (\theta - \theta_{0})^{2}$, where $k_{\theta}$ the force constant and $\theta=\cos^{-1}( \frac{\vec{r}_{ij}\cdot{\vec{r}_{jk}}}{|\vec{r}_{ij}||\vec{r}_{jk}|})$ is the angle produced by three consecutive monomers i,j,k and $\theta_{0}$ is equilibrium angle, set to 114 degrees \cite{Opaskar1965TetrahedralAI}. We choose $k=k_{\theta}$.

We place the polymeric system within two  walls:  The h-wall interact with a particle via the L-J 9-3 potential $V_{f,wh}(z_{i}) = \epsilon_{f,wh}[\frac{2}{15}(\frac{\sigma}{z_{i}})^{9} - (\frac{\sigma}{z_{i}})^{3}]$. Similarly, p-wall interact with a  particle $V_{f,wp}(z_{i}) = \epsilon_{f,wp}[\frac{2}{15}(\frac{\sigma}{z_{P}-z_{i}})^{9} - (\frac{\sigma}{z_{P}-z_{i}})^{3}]$. Here $f=(h,p)$ stands for the bead type and $z_{i}$ is the z-coordinate of i-th particle. The $\sigma$ values for all the interactions are taken to be the same but $\epsilon_{h,wh}>\epsilon_{p,wh} $ and $\epsilon_{p,wp}>\epsilon_{h,wp} $. 

The TCPs interact with each other with strength $\epsilon_{tr,tr} $ and $\epsilon_{tr,h}$ and $\epsilon_{tr,p}$ with h and p beads via L-J 12-6 potential and with two walls through L-J 9-3 potential with parameters $\epsilon_{tr,wh}$ and $\epsilon_{tr,wp}$ with h and p walls respectively. We choose them to interact more favourably with the p-beads and the p-wall than the h-beads and the h-wall. The TCPs experience force $F(z)$ which is taken to be linear in $z$ with maximum force at h-wall and vanishes at p-wall. Considering inhalation of breathing cycle, we take force over a TCP, $F(z) = F_{0}(1-z/z_{P})$. 

We take $\epsilon_{hh}$ as unit of energy. The unit of mass is the mass  of water of density 1 $gm/cm^{3}$ in a sphere of diameter $1 \mu m$. The bead diameter ($\sigma=$1 $\mu$m) is the unit of length. We set dimensionless $\epsilon_{pp}=0.33$ while the cross interaction between h and p monomers is taken to follow the Berthelot mixing rule. We set dimensionless wall particle interactions, $\epsilon_{h,wh} = 1.0$,  $\epsilon_{p,wh}=0.033$, $\epsilon_{p,wp} = 0.33$ and $\epsilon_{h,wp}=0.033$. The dimensionless L-J energy parameters for TCPs are: $\epsilon_{tr,tr}=1.0$, $\epsilon_{tr,wh} = 0.033$, $\epsilon_{tr,wp}=1.0$. Mass of the polymeric bead particles are taken to be proportional to $\sigma^{3}$ keeping the density same as water. Similarly, mass of the tracer particles are taken to be proportional to $\sigma_{tr}^{3}$ keeping the density of water. We vary several parameters,  $\bar \epsilon=\epsilon_{tr,p}/\epsilon_{tr,h}$, the network rigidity $k$ and $k_{\theta}$, and the confining length $z_{P}$. We perform simulations for various reduced temperature $T^{*} = \frac{k_{B}T}{\epsilon_{hh}}$. 

The particle dynamics are computed via the Langevin equation of motion of the ith particle  with mass $m_{i}$ at position  $\vec{r_{i}}(t)$ at time t:
\begin{equation}
m_{i} \frac{d^{2} \vec{r_{i}}}{dt ^{2}} = -\zeta_{i} \frac{d\vec{r_{i}}}{dt} - \nabla \sum_{j} V(\vec{r_{i}} - \vec{r_{j}}) + \vec{f_{i}}(t) + \vec{F_{i}}
\end{equation}
where $\zeta_{i}$ is the friction coefficient. The particle can be either the polymer beads or the TCP. $V$ is the sum of all the relevant interactions. For the network beads, we consider bonded and non-bonded interactions with other network beads, TCPs and the walls. For TCPs we consider interaction with other TCPs, network beads, the wall and in addition the external force,  $\vec{F_{i}}$. The components of $\vec{f_{i}}(t)$ are the Gaussian white noise with zero mean and variance, $6 \zeta_{i} k_{B}T \delta(t^{'}-t^{''})$, where $k_{B}$ is the Boltzmann constant, and $T$  the temperature. Here $\zeta_{i} = m_{i} \gamma$, where $\gamma$ is the damping term due to friction of the network. 

We perform simulation using the LAMMPS package \cite{LAMMPS}. Using the values of mass $m$, $\sigma$ and $\epsilon_{hh}$, we estimate a time scale, $\tau (=\sqrt{\frac{m \sigma^{2}}{\epsilon_{hh}}})\sim 0.35$ millisecond. The time step for integration is taken $0.001 \tau$. Damping term is taken as $\frac{1}{\gamma}=10 \tau$. All the quantities are averaged over five different independent trajectories each  $5000 \tau$ long. We take $L_{x}=20 \sigma$ and $L_{y}=20 \sigma$, in x- and y- directions respectively with the periodic boundary conditions (PBC) and no PBC in z-direction. We do not include hydrodynamic interaction in our calculation, which is reasonable assumption for low density system.

We create an entangled polymer network confined by h and p-wall at first. Then we insert TCPs through grand canonical Monte Carlo(GCMC) at a chemical potential and temperature corresponding to density (0.9) in bulk liquid phase near the h-wall(see  Supplementary Material, SM for details). We exert external force on TCPs and study their motion through the network medium from h-wall to p-wall. 

The network structure is characterised by the inter-strand bead-bead radial distribution function among beads, $g(r)$ \cite{allen2017computer}, calculated from the number of beads, belonging to two different strands, within a distance $r$ to $r+dr$ normalized by the corresponding shell volume and averaged over configurations. While calculating distance between two beads, we do not include periodic boundary condition in the confining z-direction. We calculate the density profile $\rho_{tr}(z)$ of the TCPs, by counting the number of tracer particles along strips of width $\delta z$ parallel to the transverse plane and dividing the number by the volume of the strip. Permeability $P(\%)$ at the p-wall is obtained by numerically integrating the density profile under the peak close to $z_{P}$ and efficiency, $e(\%)=100-P$.

 \textbf{Equilibrium Results.} -We make an entangled polymer network using the Langevin dynamics simulations as given in Eq.1 between the asymmetric walls. The network consists of 27 identical polymeric strands, each having randomly distributed 50 monomers in a number ratio h:p. An equilibrium snapshot of the entangled polymer with h:p =50:50(25 p-beads and 25 h-beads), $k=k_{\theta}=50$,  and $z_{P}=5.5$ is shown in (SM, Fig. S1 ). The $g(r)$ data in (SM Fig. S2) shows a sharp first peak at $r$=1. This confirms that the monomers among different polymeric chains stay close to each other which is signature of an entangled network. 
 
We insert TCPs above the h-wall via the GCMC at a chemical potential and temperature (SM Table S1) corresponding to the density (0.9) in bulk liquid phase. We set $\bar \epsilon = 3.3$ while keeping the other parameters same as in the entangled polymer. Fig. 1(a) shows the tracer density profile $\rho_{tr}(z)$. The data shows that the TCPs permeate through network towards the p-wall depending on $\sigma_{tr}$. Fig. 1(a)(inset) shows that $P$ drops to 4\% for $\sigma_{tr}=1.8$. This suggests that the mean pore size is ~1.8. We restrict further analysis to $\sigma_{tr} < 1.8 $.

\begin{figure}[!htb]
	\centering
	\includegraphics[width=8.0cm,height=7.3cm]{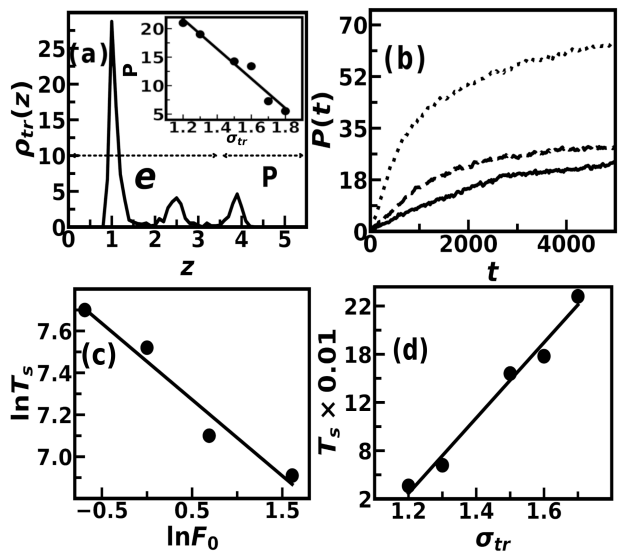}
	
	\caption{(a) Tracer density profile $\sigma_{tr}(z)$ versus $z$ plot for $T^{*}=1.0$, $\sigma_{tr}=1.5$ with $F_{0}=0$, h:p=50:50, $\bar \epsilon=3.3$, $k=k_{\theta}=50$,$z_{P}=5.5$. Inset: Permeation $P$ versus $\sigma_{tr}$ plot with $T^{*}=1.0$, $F_{0}=0$, h:p=50:50, $\bar \epsilon=3.3$, $k=k_{\theta}=50$,$z_{P}=5.5$. The solid line is the best fit. (b) $P(t)$ versus $t$ plot for  $F_{0}=1$(continuous line), $F_{0}=2$(dashed line), $F_{0}=5$(dotted line). (c) $T_{s}$ versus $F_{0}$ log-log  plot with $\sigma_{tr}=1.5$. The solid line slope -0.36 represents best fit. (d) $T_{s}$ versus $\sigma_{tr}$ plot (solid circles) with $F_{0}=1.0$. The solid line represents the best fit. The other parameters in panel (b)-(d) are the same as in (a).}

	\label{fig1 }
\end{figure}

\textbf{Permeation of tracer colloidal particles in steady state condition.} - Next we apply driving force along +z direction  on the TCPs. For $\sigma_{tr}=1 \mu m$, $<N_{tr}>=600$, $F_{0}=1$  creates pressure drop $\sim 1 $ Pa across the walls which is comparable to the experimental value\cite{konda2020aerosol} under a normal breath-ability condition. We restrict to the range, $F_{0}=1-10$.  We show in Fig. 1(b) that change in $P$ with time for $\sigma_{tr}=1.5$ for different $F_{0}$. All curves saturate at large times that ensures the steady state. We define saturation time $T_{s}$ from Fig 1.(b) when $P$ reaches 50\% of the saturation value. In Fig 1.(c) we show $T_{s}$ (for $\sigma_{tr}=1.5$) decreases with  $F_{0}$ with a power law dependence with an exponent -0.36. The $T_{s}$ increases linearly with  $\sigma_{tr}$ for $F_{0}=1.0$, shown in Fig. 1(d).  Converting to real units we find that for $\sigma_{tr}=1.5$, $T_{s}\approx 0.6\sec$. This compares well to normal healthy human breathing time. 

In order to understand the nature of permeation process, we explore the temperature dependence of $P$. We consider the saturated value of $P$ for a given condition. The  $\ln P$ versus $1/T^{*}$ plot for $\sigma_{tr}=1.5$ and different $F_{0}$ in Fig. 2(a) show linear fall, suggesting an activated Arrhenius process\cite{petrowsky2009temperature}. The slope of the linear dependence gives the activation barrier $F_{B}$. We show similar data for $\sigma_{tr}=1.2$ for $F_{0}=1.0$ in Fig. 2(a). Here the line is almost flat, denoting that the TCPs experience almost no barrier. The barrier is sensitive to  $F_{0}$ as well. We show in Fig. 2(b) that $F_{B}$ varies linearly with $F_{0}$ for two different $\sigma_{tr}=1.5$ and $1.7$. In addition, $F_{B}$ is larger for larger $\sigma_{tr}$ for a given $F_{0}$. 

Next we consider the microscopic origin of $F_{B}$. We compute the potential energy profile along the z direction per tracer particle, averaged over steady state configurations. $V_{H}(z)$ is the potential energy profile for interaction of the TCPs with the h-beads, $V_{P}(z)$ that with p-beads and the total potential energy, $V(z)=V_{H}(z)+V_{P}(z)$. We show in Fig. 2(c) $V_{H}(z),V_{P}(z)$ and $V(z)$ versus $z$ for $\sigma_{tr}=1.2$ with $F_{0}=1.0$.  We find that both $V_{H}(z)$ and $V_{P}(z)$ are almost flat, but $V_{P}(z)$ is deeper than  $V_{H}(z)$  and $V(z) \approx V_{P}(z)$. Thus, $V_{P}(z)$ is primarily responsible for the energy barrier. A larger TCP experiences more p-beads in course of its motion to experience a deeper energy landscape than smaller TCPs which we observe in Fig. 2(c) for $\sigma_{tr}=1.5$ for the same  $F_{0}$. The minima of $V_{P}(z)$ close to h-wall matches well with the $F_{B}$ value for $\sigma_{tr}=1.5$ for $F_{0}=1$ in Fig. 2(b). 

We show various factors affecting $V_{P}(z)$: (1) We plot $V_{P}(z)$ versus z for different $\bar \epsilon$ with $F_{0}=1.0$ and $\sigma_{tr}=1.5$ in Fig. 2(d). We find the minimum of the energy of the profile gets deeper at $\bar \epsilon=5$ than that at $\bar \epsilon=3.3$. For even lower $\bar \epsilon(=2.0)$, $V_{P}(z)$ is almost flat profile showing no barrier. The large tracer and p-bead interaction tends to localize the TCPs with the space of the network. (2) The localization tendency is opposed by the driving force. The external force helps the TCPs to overcome the p-bead interaction energy. We illustrate the energy profile in presence of external drive in Fig.2(d). We fix $\sigma_{tr}=1.5$ and $\bar \epsilon=3.3$ and vary $F_{0}$. A deeper minimum is observed in $V_{P}(z)$ close to the p-wall at a higher $F_{0}(=10.0)$ compared to a lower $F_{0}(=1.0)$. This shift helps the TCPs to move closer to the p-wall, increasing $P$. (3) Tracer particle localization within the mask can also be achieved by changing the network rigidity($k=k_{\theta}$). We show energy profile $V_{P}(z)$ for two different ($k=k_{\theta}$) in Fig. 4(a). We observe minimum of profile gets deeper for higher $k(=500)$ compared to lower $k(=50)$, suggesting more localisation of particles inside the network. (4) We also plot energy profile $V_{P}(z/z_{P})$ versus $z/z_{P}$ for two different confinement lengths keeping monomer density the same as in Fig. 4(b). We divide $z$ with confinement length $z_{P}$ to ensure the range (0,1) for both cases. We observe that for large $z_{P}(=16.5)$, deeper minima are observed close to the p-wall with a relatively flat profile elsewhere compared to lower $z_{P}(=5.5)$. So, particles easily passe through the network in less confined system.

\begin{figure}[!htb]
	\centering
	\includegraphics[width=8.0cm,height=7.3cm]{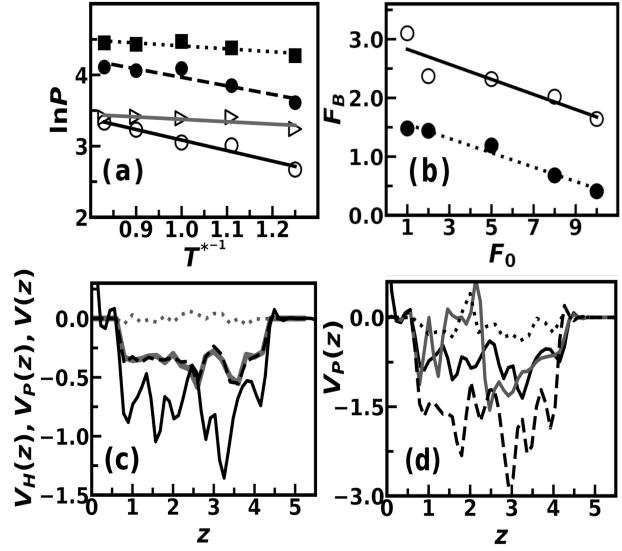}
	
	\caption{(a) $\ln P$ versus $\frac{1}{T^{*}}$ plot for $\sigma_{tr}=1.5$ with h:p=50:50,  $\bar \epsilon=3.3$, $k=k_{\theta}=50$,$z_{P}=5.5$ for $F_{0}=1$(open circle), $F=5$(close circle), $F=10$(close square) and for $\sigma_{tr}=1.2$ with $F_{0}=1$(open triangle). The best fitted lines are also shown.  (b) $F_{B}$ versus $F_{0}$ with h:p=50:50, $\bar \epsilon=3.3$, $k=k_{\theta}=50$,$z_{P}=5.5$ for $\sigma_{tr}=1.7$(open circle) and $\sigma_{tr}=1.5$(close circle) along with the fitted lines (c) Potential energy profile over $z$ for $T^{*}=1.0$, $\sigma_{tr}=1.2$ with h:p=50:50, $\bar \epsilon=3.3$, $k=k_{\theta}=50$,$z_{P}=5.5$, $F_{0}=1$: $V_{H}$(grey dotted line), $V_{P}(z)$ (grey continuous line), $V(z)$ (black dashed line). Potential Energy profile $V_{P}(z)$ (black line) for $\sigma_{tr}=1.5$ with other parameters the same. (d) $V_{P}(z)$ versus $z$ plot with $T^{*}=1.0$ for $\bar \epsilon =3.3$, $k=k_{\theta}=50$, $z_{p}=5.5$ for $F_{0}=1.0$(black line); $F_{0}=10.0$(grey line);  $\bar \epsilon =2$ (black dotted line) and $\bar \epsilon =5$ (black dashed line).}

	\label{fig2 }
\end{figure}

\begin{figure}[!htb]
	\centering
	\includegraphics[width=8.0cm,height=3.7cm]{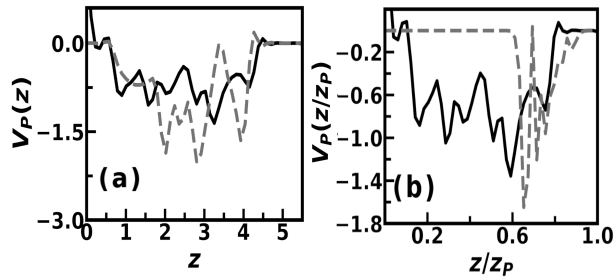}
	
	\caption{(a) $V_{P}(z)$ versus $z$ plot of $\sigma_{tr}=1.5$ with $T^{*}=1.0$, h:p=50:50, $F_{0}=1$, $\bar \epsilon=3.3$, $z_{P}=5.5$ $k=k_{\theta}=50$(black), $k=k_{\theta}=500$(dotted grey). (b) $V_{P}(z/z_{P})$ versus $z/z_{P}$ plot of $\sigma_{tr}=1.5$ with $k=k_{\theta}=50$ for $z_{P}=5.5$ (black), $z_{P}=16.5$ (dotted grey). Other parameters are same as panel (a). }
	\label{fig3 }
\end{figure}  

\textbf{Mask efficiency.} - Next we illustrate how the mask efficiency $e$ can be tuned by the network properties. We fix the temperature at the room temperature  $(T^{*}=1.0)$, make sure of the breath-ability condition $(F_{0}=1.0)$ and consider the droplet size in the micrometer range $(\sigma_{tr}=1.5)$. We observe changes in $e$ varying $\bar \epsilon$ and h:p ratio. We show  $e$ versus $\bar \epsilon $ plots in Fig.4(a) with three different h:p ratio. Here $e$ increases with $\bar \epsilon$ which is consistent, showing two distinct regions of linear dependence for h:p=50:50, with slope 2.0 for low $\bar \epsilon$, followed by much slower increase with slope 0.3. The increase in $e$ is consistent with the change in the energy barrier with $\bar \epsilon$. The crossover in the dependence results from increasing population of TCPs within the network. For h:p=30:70  there is similar change in the slopes, although the $e$ values are less compared to the corresponding case for h:p=50:50. The decrease is observed despite having large number of p-beads than the 50:50 case. For h:p=70:30, $e$ shows a slow change with  $\bar \epsilon$ compared to the other cases. Physically, the composition dependence can be understood as follows. As the number of h-beads decreases, the network gets loose. A loose network performs less efficiently. This can be observed from $g(r)$ data for three different cases in Fig.4(b). A larger first peak for 70:30 case indicates a stronger network connectivity which is observed with large number of h-beads. However, for 70:30 case even if the network structure is stronger, the p-beads are less abundant, reducing the efficiency.

\begin{figure}[!htb]
	\centering
	\includegraphics[width=8.0cm,height=3.7cm]{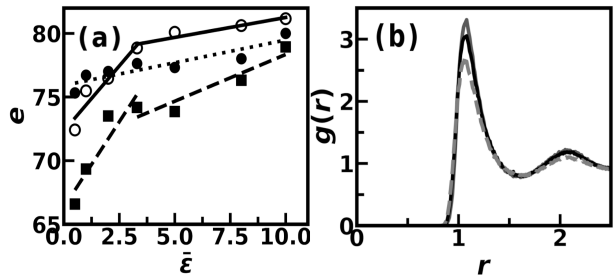}
	
	\caption{(a) $e$ versus $\bar \epsilon$ plots of $\sigma_{tr}=1.5 $ with $T^{*}=1.0$, $k=k_{\theta}=50$,$z_{P}=5.5$,$F_{0}=1$, h:p=50:50 (open circle, the solid line being best fits) , h:p=30:70(close square, the dashed line being best fit) , h:p=70:30(close circle, the dotted line is best fit with single slope 0.35). (b) pair correlation function of beads belonging to two different polymeric strands $g(r) $ over r for h:p-50:50(black) , h:p-70:30(grey), h:p-30:70(dotted grey) with $T^{*}=1.0$, $\epsilon_{hh} = 1.0$, $\epsilon_{pp} = 0.33$, $\epsilon_{h,wh} = 1.0$, $\epsilon_{p,wh}=0.033$, $\epsilon_{p,wp} = 0.33$, $\epsilon_{h,wp}=0.033$, $k=k_{\theta}=50$,$z_{P}=5.5$.}
	\label{fig4 }
\end{figure}

We also examine how the rigidity and confinement length $z_{P}$ affect $e$. We tune the strength of the bonded  interactions $k(=k_{\theta})$ and plot $e$ with $k$ in Fig. 5(a). We observe  linear increase in $e$ with $k$ and saturation for large $k$. Unlike the case of Fig. 4(b) there is no such significant change in the entanglement property of the network for different rigidities as shown by $g(r)$ for different $k$ values in the inset of Fig. 5(a). We further compute the normal (zz) component of the pressure tensor (see details in SM) of the network beads as a function of z. The normal pressure profile $P_{zz}(z)$ in Fig. 5(b) shows peaks close to both walls. The peak of $P_{zz}(z)$ close to p-wall increases with rigidity.  Thus, the network resists more particles from passing through it towards p-wall with increasing $k$ resulting in enhanced efficiency. This is consistent with deeper $V_{P}(z)$ in Fig. 3(a) observed for large $k$. In other words, a large $k$ allows a better interaction of the network with the TCP. In order to examine how the thickness of the network region affects $e$, we increase  $z_{P}$ keeping the monomer density fixed. We plot $e$ with $z_{P}$ in Fig. 6(a). We find as $z_{P}$ increases as $e$  decreases linearly and eventually gets saturation beyond $z_{P}=16.5$. Fig. 6(b) shows that the first peak height of $g(r)$ (SM, Fig. S3), $g_{max}$ linearly drops with $z_{P}$ up to $z_{P}=16.5$ after which the bulk limit (see details in SM) sets in. Thus, the system gets less confined, the network structure gets loose. A loose network provides a flat $V_{P}(z)$, shown in Fig. 3(b), and better mobilities of the TCPs and hence, poorer $e$.

 \begin{figure}[!htb]
	\centering
	\includegraphics[width=8.0cm,height=3.7cm]{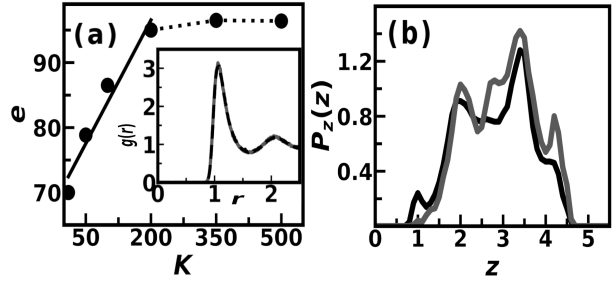}
	
	\caption{(a) $e$ versus $k$ plot (solid circles) for $\sigma_{tr}=1.5$ with $T^{*}=1.0$, h:p=50:50, $F_{0}=1$, $\bar \epsilon=3.3$, $z_{P}=5.5$. Solid line  represents the best fit and dotted line is a guide to the eyes). Inset: pair correlation function of beads belonging to two different polymeric strands $g(r)$ versus $r$ for $k=k_{\theta}=50$(dotted black) and $k=k_{\theta}=500$(grey). Other parameters are $T^{*}=1.0$,h:p-50:50, $\epsilon_{hh} = 1.0$, $\epsilon_{pp} = 0.33$, $\epsilon_{h,wh} = 1.0$, $\epsilon_{p,wh}=0.033$, $\epsilon_{p,wp} = 0.33$, $\epsilon_{h,wp}=0.033$, $z_{P}=5.5$. (b) $P_{z}(z)$ versus $z$ plots for $k=k_{\theta}=50$(black) and $k=k_{\theta}=500$(grey). Other parameters are as same as panel (a).}

	\label{fig5 }
\end{figure}

\begin{figure}[!htb]
	\centering
	\includegraphics[width=5.6cm,height=13.7cm]{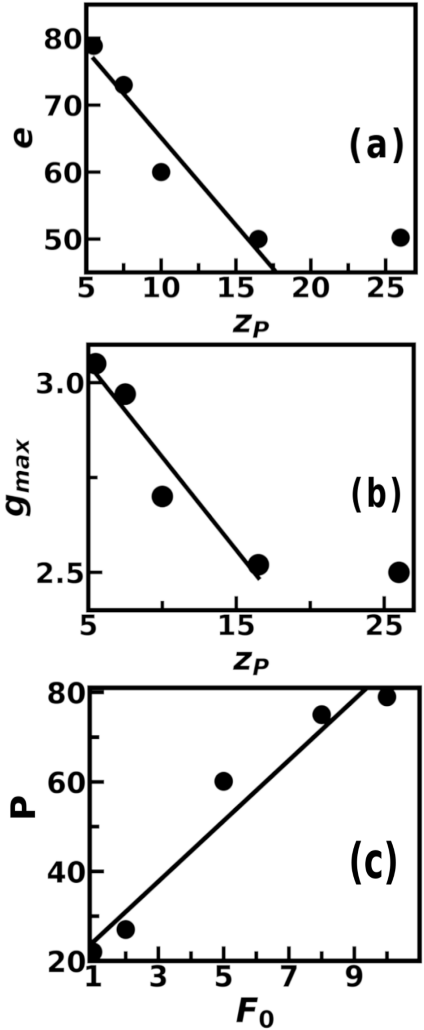}
	
	\caption{(a) $e$ versus $z_{P}$ plot (solid circles) for $T^{*}=1.0$, $\sigma_{tr}=1.5$ with $F_{0}=1$, h:p=50:50, $\bar \epsilon=3.3$, $k=k_{\theta}=50$. (b) $g_{max}$ versus $z_{P}$ (solid circles). Parameters are $T^{*}=1.0$, h:p-50:50, $\epsilon_{hh} = 1.0$, $\epsilon_{pp} = 0.33$, $\epsilon_{h,wh} = 1.0$, $\epsilon_{p,wh}=0.033$, $\epsilon_{p,wp} = 0.33$, $\epsilon_{h,wp}=0.033$, $k=k_{\theta}=50$. (c) $P$ versus $F_{0}$ plot(solid circles) for the reverse case with $T^{*}=1.0$, h:p=50:50, $\bar \epsilon=3.3$, $k=k_{\theta}=200$, $z_{P}=5.5$. Solid line represents the best fits.}

	\label{fig6 }
\end{figure}

Our results suggest that large efficiency can be achieved by a deeper $V_{P}(z)$ profile. For instance, $95\%$ efficiency in intercepting tracer micro droplets under normal breath-ability condition at room temperature can be achieved under the following conditions: (1) Composition of h:p=50:50 so that the number of p-beads is sufficient in number to interact with the TCPs, keeping the network structure sufficiently robust. (2) $\bar \epsilon > 3$ to allow large interaction of the p-beads with the TCPs. (3) Fairly rigid network ($k \ge 200$); and (4) $z_{P} \sim$ few microns so that the network remains sufficiently strong to resist the TCP penetration. So far we have calculated tracer permeation during inhalation cycle. Now we consider exhalation with FM on. Using  parameters for $e \sim 95 \%$ ($\bar \epsilon=3.3$, $F_{0}=1.0$, $T^{*}=1.0$, $k=200$, and $z_{P}=5.5$) we put TCPs with $\sigma_{tr}=1.5$ close to p-wall as done previously. We apply driving force from p-wall to h-wall. We find in Fig 6.(c) that the $P$ at h-wall increases linearly with $F_{0}$. Thus, TCPs are efficiently drained away. 

\textbf{Conclusions.}- To summarize,  we study the permeation of TCPs within a network of polymer in presence of a driving force, using Langevin dynamics simulations. The parameters in our studies are chosen so as to ensure breath-ability conditions in an FM. We find that the droplet permeation through the network is an activated process. The activation barrier increases with (1) increasing tracer size, (2) decreasing driving force and (3) increasing network-tracer interactions. The barrier height is mainly determined by the magnitude of the more favourable interaction experienced by the TCP in the network. For a given tracer size, driving force and temperature, the efficiency of the mask depends on: (1) h:p ratio, (2) increasing interaction strength of tracer and p-beads, (3) increasing the rigidity and (4) decreasing the thickness of the network. It may be noted that efficiency of the model mask does not depend on electrostatic interaction. Our model is sufficiently robust to be helpful in designing mask with better efficiency while retaining normal breath-ability conditions.

\textbf{Acknowledgements}
R.K thanks DST Inspire for financial support. The authors thank the Thematic Unit of Excellence(TUE) and the Technical Research Centre (TRC) at S.N.Bose National Centre for Basic Sciences for computational facilities.

\section*{ASSOCIATED CONTENT}
 Supplementary Material contains a brief description of (a)  Insertion of tracer colloidal particles, (b) Pressure profile of network beads, (c) Entanglement without confinement. Table S1 tabulates chemical potential values and
Fig. S1, S2 and S3.

\bibliographystyle{vancouver}
\bibliography{paper.bib}{}

\newpage

\newpage

\end{document}